\documentstyle[12pt]{article}

\begin{document}
\setcounter{page}{1}
\pagestyle{plain} \vspace{1cm}
\begin{center}
\Large{\bf  Some Aspects of Planck Scale Quantum Optics}\\
\small
\vspace{1cm}
{\bf Kourosh Nozari}\\
\vspace{0.5cm}
{\it Department of Physics,
Faculty of Basic Science,\\
University of Mazandaran,\\
P. O. Box 47416-1467,
Babolsar, IRAN\\
e-mail: knozari@umz.ac.ir}
\end{center}
\vspace{1.5cm}
\begin{abstract}
This paper considers the effects of gravitational induced
uncertainty on some well-known quantum optics issues. First we will
show that gravitational effects at quantum level destroy the notion
of harmonic oscillations. Then it will be shown that, although it is
possible(at least in principle) to have complete coherency and
vanishing broadening in usual quantum optics, gravitational induced
uncertainty destroys complete coherency and it is impossible to have
a monochromatic ray. We will show that there is an additional
wave packet broadening due to quantum gravitational effects.
\end{abstract}
\newpage

\section{Introduction}
Harmonic analysis is a primary input for a vast number of technics
and approaches in quantum optics. The possible break down of this
simple notion which is the essence of Fourier analysis, should
result in a variety of novel implications. If one be able to show
that there is no harmonic oscillation essentially, a number of
technics and concepts should be re-examined. Here, we will show
that, when one considers quantum effects of gravity, the very notion
of harmonicity breaks down. This feature implies some new
implications for the rest of quantum optics. In usual quantum
optics, one can have coherent states in principle. These states are
states with minimum uncertainties(maximum localization) and
therefore minimum broadening when they propagates. In other words,
based on Heisenberg uncertainty principle, $\Delta x\Delta
p\geq\hbar$, it is possible, in principle, to have localized states,
and as a result, a wave packet can propagates from one point to
another point without any broadening(the so-called solitonic
states). When one considers gravitational effect at quantum level,
the situation differs considerably. Gravity induces uncertainty and
this extra uncertainty will produce new quantum optical phenomena.
As a result, although it is possible to have complete coherency and
vanishing broadening in usual quantum mechanics, gravitational
induced uncertainty destroys complete coherency and it is not
possible to have a monochromatic ray in principle. The goal of this
paper is the investigation of such a new quantum gravitational
induced phenomena.\\
The structure of the paper is as follows: Section 2 gives an
overview to Generalized Uncertainty Principle(GUP). In section 3 we
will show that there is no harmonic oscillation in gravitational
quantum optics. In section 4 the problem of coherent states for
harmonic oscillation is discussed. We will show that due to the
failure of the notion of harmonic oscillation, although there is no
considerable difference in definition of coherent states relative to
ordinary quantum mechanics, considering expectation values and
variance of some operators, quantum gravitational arguments leads to
the result that complete coherency is impossible in extreme quantum
gravity regime. Section 5 considers the effect of gravitation on
wave packet propagation. We will show that there is an extra
broadening due to gravitational induced uncertainty. Summary and
conclusions are presented in section 6.
\section{Generalized Uncertainty Principle }
Recently it has been indicated that measurements in quantum gravity
should be governed by generalized uncertainty principle. There are
some evidences from string theory[1-5], black holes Physics gedanken
experiments[6,7] and loop quantum gravity[8], which leads some
authors to re-examine usual uncertainty principle of Heisenberg.
These evidences have origin on the quantum fluctuation of the
background spacetime metric. Introduction of this idea has drown
considerable attention and many authors considered various problems
in the framework of generalized uncertainty principle[9-20]. Such
investigations have revealed that in Planck scale a re-formulation
of quantum theory is un-avoidable. This re-formulated quantum theory
should incorporate gravitational effects from very beginning. In
this extreme quantum level, spacetime is not commutative[21] and
based on some general arguments it is possible to interpret gravity
as a consequence of some unknown quantum effects[22]. As another
novel consequence of such re-formulated quantum theory, constants of
the nature may vary with time[23,24]. In addition, the very notion
of locality and position space representation are not satisfied in
Planck scale[25,26] and one has to consider Hilbert space of
maximally localized states. In the same manner which uncertainty
principle provides a thorough foundation for usual quantum theory,
now generalized uncertainty principle(GUP) is the cornerstone of
modified quantum theory. Generalized uncertainty principle leads
naturally to the existence of a minimal observable length on the
order of Plank length $l_{P}$. A generalized uncertainty principle
can be formulated as
\begin{equation}
\label{math:2.1}\Delta x\geq\frac{\hbar}{\Delta p}+const.G\Delta
p,
\end{equation}
which, using the minimal nature of $ l_{P} $ can be written as,
\begin{equation}
\label{math:2.2} \Delta x\geq\frac{\hbar}{\Delta
p}+\alpha^{\prime}l_{p}^2 \frac{\Delta p}{\hbar}.
\end{equation}
The corresponding Heisenberg commutator now becomes,
\begin{equation}
\label{math:2.3} [x,p]=i\hbar(1+\beta p^2).
\end{equation}
Actually as Kempf {\it et al} have argued[25], one can consider more
generalization such as
\begin{equation}
\label{math:2.4} \Delta x\Delta
p\geq\frac{\hbar}{2}\Big(1+\alpha(\Delta x)^2+\beta(\Delta
p)^2+\gamma \Big)
\end{equation}
and the corresponding commutator relation is
\begin{equation}
\label{math:2.5}[x,p]=i\hbar(1+\alpha x^2+\beta p^2).
\end{equation}
This statement shows that GUP itself has a perturbational expansion.
In which follows, since we are dealing with dynamics, we consider
only equation (2) or equivalently (3). The main consequence of this
GUP is that measurement of position is possible only up to Plank
length, $ l_{P} $. So one can not setup a measurement to find more
accurate particle position than Plank length. In other words, one
can not probe distances less than Planck length.

\section{GUP and Harmonic Oscillations}
The problem of harmonic oscillation in the context of GUP first has
been considered by Kempf {\it et al}[25]. They have found
eigenvalues and eigenfunctions of harmonic oscillator in the context
of GUP by direct solving of the Schr$\ddot{o}$dinger equation. Then
Camacho has analyzed the role that GUP can play in the quantization
of electromagnetic field. He has considered electromagnetic
oscillation modes as simple harmonic oscillations[11,15]. Here we
proceed one more step to find dynamics of harmonic oscillator in the
framework of GUP using Heisenberg picture of quantum mechanics. In
Heisenberg picture of quantum mechanics, equation of motion for
observable $A$ is as follows,
\begin{equation}
\label{math:3.1}\frac{dA}{dt}= \frac{i}{\hbar}[H,A].
\end{equation}
Hamiltonian for a simple harmonic oscillator is,
\begin{equation}
\label{math:3.2}H=\frac{p^2}{2m}+ \frac{1}{2}m\omega^{2} x^2
\end{equation}
Now the equations of motion for $x$ and $p$ are respectively,
\begin{equation}
\label{math:3.3}\frac{dx}{dt}= \frac{1}{m}\Big(p+\beta p^3\Big),
\end{equation}
and
\begin{equation}
\label{math:3.4}\frac{dp}{dt}=
-\frac{1}{2}m\omega^{2}\Big(2x+\beta x p^2 +\beta p^2 x\Big).
\end{equation}
Using Baker-Hausdorff lemma, a lengthy calculation gives the
following equations for time evolution of $x$ and $p$
respectively,

$$x(t) = x(0)\cos\omega t+ \frac{p(0)}{m\omega}
\sin \omega t$$
$$+ \beta \Bigg[\frac{p^{3}(0)}{m\omega}(\omega t) -
\frac{1}{2}\Bigg(p(0)x(0)p(0)+\frac{3}{2}\Big[x(0)p^{2}(0)+
p^{2}(0)x(0)\Big] \Bigg)(\omega t)^2$$
 $$- \Bigg(\frac{5}{6}
\frac{p^{3}(0)}{m\omega}-\frac{5}{12} m\omega\Big[x^{2}(0)p(0)+
p(0)x^{2}(0)\Big] - \frac{1}{2}m\omega x(0)p(0)x(0)\Bigg)(\omega
t)^3$$
\begin{equation}
\label{math:3.5} + \Bigg(\frac{11}{24} \Big[x(0)p^{2}(0) +
p^{2}(0) x(0)\Big] +
\frac{5}{12}p(0)x(0)p(0)-\frac{1}{3}m^{2}\omega^{2} x^{3}(0)\Bigg)
(\omega t )^4 \Bigg],
\end{equation}
and
$$p(t) = p(0)\cos\omega t- m\omega x(0)\sin \omega t$$
$$+ \beta \Bigg[-\frac{1}{2}m\omega \Big[x(0)p^{2}(0)+
p^{2}(0)x(0)\Big](\omega t)$$
$$-\Bigg( p^{3}(0) -\frac{1}{4}
m^{2}\omega^{2}\Big[ p(0)x^{2}(0)+ x^{2}(0)p(0)
+2x(0)p(0)x(0)\Big]\Bigg)(\omega t)^2$$
\begin{equation}
\label{math:3.5} +\Bigg(\frac{2}{3} m\omega\Big[x(0)p^{2}(0)+
p^{2}(0)x(0)\Big]+ \frac{1}{2}p(0)x(0)p(0)-\frac{1}{3}
m^{3}\omega^{3} x^{3}(0)\Bigg)(\omega t)^{3}\Bigg],
\end{equation}
where only terms proportional to first order of $\beta$ are
considered. It is evident that in the limit of $\beta\rightarrow 0$
one recover the usual results of ordinary quantum mechanics. The
term proportional to $\beta$ shows that in the framework of GUP
harmonic oscillator is no longer "harmonic" essentially, since, now
its time evolution has not oscillatory nature completely. In other
words, in the framework of GUP there is no harmonic motion and this
is a consequence of gravitational effect at quantum level.\\
Now for computing expectation values, we need a well-defined
physical state. Note that eigenstates of position operators are not
physical states because of existence of a minimal length which
completely destroys the notion of locality. So we should consider a
physical state such as $|\alpha\rangle$ where $|\alpha\rangle$ is
for example a maximally localized or momentum space eigenstate[25].
Suppose that $p_{\alpha}(0)=\langle\alpha|p(0)|\alpha\rangle$ and
$x_{\alpha}(0)=\langle\alpha|x(0)|\alpha\rangle$. Now the
expectation value of momentum operator is,
$$\frac{\langle\alpha|p(t)|\alpha\rangle}{m} = \frac{p_{\alpha}(0)}{m}\cos\omega t- \omega x_{\alpha}(0)\sin \omega t$$
$$+ \beta \Bigg[-\frac{1}{2}\omega \Big(x_{\alpha}(0)p_{\alpha}^{2}(0)+
p_{\alpha}^{2}(0)x_{\alpha}(0)\Big)(\omega t)$$
$$-\Bigg( \frac{p_{\alpha}^{3}(0)}{m} -\frac{1}{4}
m\omega^{2}\Big[ p_{\alpha}(0)x_{\alpha}^{2}(0)+
x_{\alpha}^{2}(0)p_{\alpha}(0)
+2x_{\alpha}(0)p_{\alpha}(0)x_{\alpha}(0)\Big]\Bigg)(\omega t)^2$$
\begin{equation}
\label{math:3.6} +\Bigg(\frac{2}{3}
\omega\Big[x_{\alpha}(0)p_{\alpha}^{2}(0)+
p_{\alpha}^{2}(0)x_{\alpha}(0)\Big]+
\frac{1}{2m}p_{\alpha}(0)x_{\alpha}(0)p_{\alpha}(0)-\frac{1}{3}
m^{2}\omega^{3} x_{\alpha}^{3}(0)\Bigg)(\omega t)^{3}\Bigg].
\end{equation}
This relation shows that there is a complicated dependence of the
expectation value of momentum operator to the mass of the
oscillator. In usual quantum mechanics,
$\frac{\langle\alpha|p(t)|\alpha\rangle}{m}$ and $
\frac{p_{\alpha}(0)}{m}$ are mass independent. Here although $
\frac{p_{\alpha}(0)}{m}$ is still mass independent, but now
$\frac{\langle\alpha|p(t)|\alpha\rangle}{m}$ has a complicated mass
dependence. This is a novel implication which have been induced by
GUP. Physically, it is completely reasonable that the expectation
value for momentum of a particle be a function of its mass, but the
mass dependence here has a complicated form relative to usual
situation.
\section{Gup and Coherency}
As a consequence of gravitational induced uncertainty, it seems that
some basic notions such as coherency should be re-examined in this
new framework. Here we want to show that in quantum gravity regime
there is no coherent state at all. We consider the simple harmonic
oscillator by Hamiltonian
\begin{equation}
\label{math:2.6} H=\frac{1}{2m}(p^2+m^2\omega^2x^2)
\end{equation}
The problem of quantum oscillator is easily solved in terms of the
annihilation and creation operators $a$ and $a^{\dagger}$. We
recall the fundamental definitions:
\begin{equation}
\label{math:2.7}
a=\sqrt{\frac{m\omega}{2\hbar}}\big(x+\frac{ip}{m\omega}\big),
\end{equation}
\begin{equation}
\label{math:2.7}a^{\dagger}=\sqrt{\frac{m\omega}{2\hbar}}\big(x-\frac{ip}{m\omega}\big)
\end{equation}
and the inverse relations:
\begin{equation}
\label{math:2.8}x=\sqrt{\frac{\hbar}{2m\omega}}(a+a^{\dagger}),\qquad
p=i\sqrt{\frac{m\hbar\omega}{2}} (-a+a^{\dagger}).
\end{equation}
The Hamiltonian H is given in terms of these operators as :
\begin{equation}
\label{math:2.9}H=\hbar\omega(a^{\dagger}a+\frac{1}{2})
\end{equation}
If we set $ N\equiv a^{\dagger} a $ (: Number operator),then
\begin{equation}
\label{math:2.10}[N,a^{\dagger}]=a^{\dagger}, \quad
[N,a]=-a,\qquad [a^{\dagger},a]=-1
\end{equation}
Let $\textbf{H}$ be a Fock space generated by $a$ and $
a^{\dagger}$, and $\{|n\rangle| n\in \textbf\{N\}\cup\{0\}\}$ be
its basis. The action of $a$ and $a^{\dagger}$ on $\textbf{H}$ are
given by
\begin{equation}
\label{math:2.11}a|n\rangle=\sqrt{n}|n-1\rangle, \qquad
a^{\dagger}|n\rangle=\sqrt{n+1}|n+1\rangle, \qquad
N|n\rangle=n|n\rangle
\end{equation}
Where $|0\rangle$ is a normalized vacuum ($a|0\rangle = 0$  and
$\langle0|0\rangle = 1$). Therefore states $|n\rangle$ for $ n\geq1$
are given by
\begin{equation}
|n\rangle=\frac{{a^{\dagger}}^n}{\sqrt{n!}}|0\rangle.
\end{equation}
These states satisfy the orthogonality and completeness conditions
\begin{equation}
\label{math:2.12}\langle m|n\rangle =\delta_{mn}, \qquad
\sum_{n=0}^{\infty}|n\rangle\langle n|=1.
\end{equation}
The coherent state was introduced by Schr\"{o}dinger as the quantum
state of the harmonic oscillator which minimizes the uncertainty
equally distributed in both position $x$ and momentum $p$. By
definition, coherent state is the normalized state $|\lambda\rangle
\in \textbf{H}$, which is the eigenstate of annihilation operator
and satisfies the following equation,
\begin{equation}
 \label{math:2.13}a|\lambda\rangle =\lambda|\lambda\rangle \qquad where \qquad \langle\lambda|\lambda\rangle=1
\end{equation}
and
\begin{equation}
\label{math:2.13}|\lambda\rangle=e^{-|\lambda|^2/2}\sum_{n=0}^{\infty}\frac{\lambda^n}
{\sqrt{n!}}|n\rangle=e^{-|\lambda|^2/2}e^{\lambda
a^{\dagger}}|0\rangle.
\end{equation}
Actually $\lambda$ can be complex because $a$ is not Hermittian. Let
us now consider the following possibility, as a generalization for
creation and annihilation operators in GUP,
\begin{equation}
\label{math:3.1}  a=\frac{1}{\sqrt{2\hbar\omega}}\bigg(\omega
x+i[p+f(p)]\bigg),
\end{equation}
\begin{equation}
\label{math:3.2}
a^{\dagger}=\frac{1}{\sqrt{2\hbar\omega}}\bigg(\omega
x-i[p+f(p)]\bigg).
\end{equation}
Here $f(p)$ is a function that satisfies three conditions, namely:
(i) in the limit $\beta\rightarrow0 $ we recover the usual
definition for the creation and annihilation operators,(14) and
(15); (ii) if $\beta\neq0$, then we have (3), and; (iii)
$[a_{\vec{k}},a^{\dagger}_{\vec{k^{\prime}}}]=i\hbar
\delta_{\vec{k}\vec{k^{\prime}}}$, where $\vec{k}$ and
$\vec{k^{\prime}}$ are corresponding wave vectors. It can be shown
that the following function satisfies the aforementioned
restrictions
\begin{equation}
\label{math:3.3}
f(p_{\vec{k}})=\sum_{n=1}^{\infty}\frac{(-\beta)^n}{2n+1}p_{\vec{k}}^{2n+1}
\end{equation}
Condition (iii) means that the usual results, in relation with the
structure of the Fock space, are valid in our case, for instance,
the definition of the occupation number operator, $N_{\vec{k}}=
a^{\dagger}_{\vec{k}} a_{\vec{k}}$ , the interpretation of
$a^{\dagger}_{\vec{k}}$ and $a_{\vec{k}}$ are creation and
annihilation operators, respectively, etc. Clearly, the relation
between $p_{\vec{k}}$, $a_{\vec{k}}$ and $a^{\dagger}_{\vec{k}}$ is
not linear, and from the Hamiltonian (13) we now deduce that it is
not diagonal in the occupation number representation. Let us now
consider
\begin{equation}
\label{math:3.4} f(p_{\vec{k}})=-\frac{\beta}{3}p_{\vec{k}}^3
\end{equation}
In this form we find $p_{\vec{k}}$ as a function of $a_{\vec{k}}$
and $a^{\dagger}_{\vec{k}}$, namely
\begin{equation}
\label{math:3.5}
p_{\vec{k}}=-i\sqrt{\frac{\hbar\omega}{2}}\big(a_{\vec{k}}-a^{\dagger}_{\vec{k}}\big)
\big[1-\sqrt{\frac{\hbar\omega\beta}{8}}(a_{\vec{k}}-a^{\dagger}_{\vec{k}})\big]
\end{equation}
It is clear that, if $\beta=0$ we recover the usual case.
Rephrasing the Hamiltonian as a function of the creation and
annihilation operators we find:
\begin{equation}
\label{math:3.6}
H=\sum_{\vec{k}}\hbar\omega\big[N_{\vec{k}}+\sqrt{\frac{\hbar\omega\beta}{8}}
g(a_{\vec{k}},a^{\dagger}_{\vec{k}})+\beta\frac{(\hbar\omega)^2}{16}h(a_{\vec{k}},a^{\dagger}_{\vec{k}})\big]
\end{equation}
where functions $ g(a_{\vec{k}},a^{\dagger}_{\vec{k}})$ and
$h(a_{\vec{k}},a^{\dagger}_{\vec{k}}) $ are:
\begin{equation}
\label{math:3.7}g(a_{\vec{k}},a^{\dagger}_{\vec{k}})=a_{\vec{k}}^3-N_{\vec{k}}a_{\vec{k}}
-a_{\vec{k}}N_{\vec{k}}-a_{\vec{k}}-(a^{\dagger}_{\vec{k}})^3+N_{\vec{k}}
a^{\dagger}_{\vec{k}}+a^{\dagger}_{\vec{k}}N_{\vec{k}}+a^{\dagger}_{\vec{k}}
\end{equation}
and
\begin{equation}
\label{math:3.8}
    h(a_{\vec{k}},a^{\dagger}_{\vec{k}})=a_{\vec{k}}^4+
 a_{\vec{k}}^2(a^{\dagger}_{\vec{k}})^2-a_{\vec{k}}^3a^{\dagger}_{\vec{k}}-
 a_{\vec{k}}^2a^{\dagger}_{\vec{k}}a_{\vec{k}} $$ $$ +(a^{\dagger}_{\vec{k}})^2a_{\vec{k}}^2+
 (a^{\dagger}_{\vec{k}})^4-
(a^{\dagger}_{\vec{k}})^2a_{\vec{k}}a^{\dagger}_{\vec{k}}-
(a^{\dagger}_{\vec{k}})^3a_{\vec{k}}$$
$$ -a_{\vec{k}}a^{\dagger}_{\vec{k}}a_{\vec{k}}^2-
a_{\vec{k}}(a^{\dagger}_{\vec{k}})^3+a_{\vec{k}}a^{\dagger}_{\vec{k}}a_{\vec{k}}a^{\dagger}_{\vec{k}}
+a_{\vec{k}}(a^{\dagger}_{\vec{k}})^2a_{\vec{k}} $$
$$ -a^{\dagger}_{\vec{k}}a_{\vec{k}}^3-a^{\dagger}_{\vec{k}}a_{\vec{k}}(a^{\dagger}_{\vec{k}})^2+
a^{\dagger}_{\vec{k}}a_{\vec{k}}^2a^{\dagger}_{\vec{k}}+
a^{\dagger}_{\vec{k}}a_{\vec{k}}a^{\dagger}_{\vec{k}}a_{\vec{k}}.
\end{equation}
Now with these pre-requisites we can consider the coherent states
in the context of GUP. Suppose $|\lambda\rangle$ be an eigenstate
of the annihilation operator. We remember that the definition of
the annihilation operator in GUP may be different from the usual
quantum mechanics but the fact that eigenstates of annihilation
operator are coherent states do not changes. Therefore one can
write
\begin{equation}
\label{math:3.9}   a|\lambda\rangle=\lambda|\lambda\rangle
\end{equation}
Indeed $|n\rangle$ is the eigenstate of the number operator and
satisfies completeness and orthogonality conditions. So we can
expand $|\lambda\rangle$ in terms of the stationary states
$|n\rangle$
\begin{equation}
\label{math:3.10}
|\lambda\rangle=\sum_{n=0}^{\infty}|n\rangle\langle
n|\lambda\rangle=C_{n}|n\rangle,
\end{equation}
The eigenvalue equation (19) implies the following recursion formula
for the expansion coefficients:
\begin{equation}
\label{math:3.11}C_{n}=\frac{\lambda}{\sqrt{n}}C_{n-1}.
\end{equation}
We immediately obtain
\begin{equation}
\label{math:3.12}C_{n}=\frac{\lambda^{n}}{\sqrt{n!}} C_{0},
\end{equation}
The constant $ C_{0}$ is determined from the normalization
condition on the Fock space,
\begin{equation}
\label{math:3.13}1=\langle\lambda|\lambda\rangle
=|C_{0}|^2\sum_{n=0}^{\infty}\frac{\lambda^{2n}}{\sqrt{n!}}=|C_{0}|^2e^{|\lambda|^2},
\end{equation}
For any complex number $\lambda$ the correctly normalized
quasi-classical state $|\lambda\rangle$ is therefore given by
\begin{equation}
\label{math:3.14}|\lambda\rangle=e^{-\frac{1}{2}|\lambda|^2}\sum\frac{|\lambda|^n}{\sqrt{n!}}|n\rangle.
\end{equation}
We recall that the n-th stationary state $|n\rangle$ is obtained
from the ground state wave function by repeated application of the
operator $a^{\dagger}$,
\begin{equation}
\label{math:3.15}|n\rangle=\frac{1}{\sqrt{n!}}(a^{\dagger})^n|0\rangle,
\end{equation}
This allows us to write the coherent state in the form:
\begin{equation}
\label{math:3.16}|\lambda\rangle=e^{-\frac{1}{2}|\lambda|^2}\sum_{n=0}^{\infty}\frac{1}{n!}(\lambda
a^{\dagger})^n|0\rangle = e^{-\frac{1}{2}|\lambda|^2} e^{\lambda
a^{\dagger}}|0\rangle
\end{equation}
We see that this expression for the eigenstates of the annihilation
operator is the same as usual quantum mechanics, equation (23).
Actually, it is not surprising that there is no changes in the form
of states by modifying the uncertainty relation and similarly for
the coherent state. The unchanged state itself cannot be the result
of considering generalized uncertainty principle (GUP). It is
because a quantum state does not necessarily imply a direct
connection with uncertainty principle. Differences caused by
different uncertainty relations (such as the GUP) will be found in
the expectation values of the operators for a given state and their
statistics(such as variance) that can be obtained from the
measurement on the state. To analyze the coherent state under the
GUP, we should consider $\langle x\rangle$ and $\langle p\rangle$
for the coherent state and see that whether they are changed or not.
For this end, suppose that $|\lambda \rangle$ is a coherent state
given by (39). Since
\begin{equation}
\label{math:5.1}x=\sqrt{\frac{\hbar}{2\omega}}(a_{\vec{k}}
+a^{\dagger}_{\vec{k}}),
\end{equation}
and
\begin{equation}
\label{math:5.2}p=-i\sqrt{\frac{\hbar\omega}{2}}(a_{\vec{k}}-a^{\dagger}_{\vec{k}})
[1-\sqrt{\frac{\hbar\omega\beta}{8}}(a_{\vec{k}}-a^{\dagger}_{\vec{k}})],
\end{equation}
one finds the following result for the expectation value of position
operator, $x$
\begin{equation}
\label{math:5.3}\langle
x\rangle=\langle\lambda|x|\lambda\rangle=\sqrt{\frac{\hbar}{2\omega}}
\langle\lambda|a_{\vec{k}}+a^{\dagger}_{\vec{k}}|\lambda\rangle=
\sqrt{\frac{\hbar}{2\omega}}(\lambda+\lambda^{\ast}).
\end{equation}
Therefore one has,
\begin{equation}
\label{math:5.4} \langle
x\rangle^2=\frac{\hbar}{2\omega}(\lambda^2+{\lambda^{\ast}}^2
+2\lambda\lambda^{\ast})=\frac{\hbar}{2\omega}(\lambda+\lambda^{\ast})^2.
\end{equation}
It is straightforward to show that,
\begin{equation}
\label{math:5.5}\langle
x^2\rangle={\frac{\hbar}{2\omega}}(\lambda^2+{\lambda^{\ast}}^2+
2\lambda\lambda^{\ast}+1)=\frac{\hbar}{2\omega}(\lambda+\lambda^{\ast})^2+1,
\end{equation}
and therefore we find for the variance of $x$,
\begin{equation}
\label{math:5.6}(\Delta x)^2=\langle x^2\rangle-\langle
x\rangle^2=\frac{\hbar}{2\omega}.
\end{equation}
This is the same as usual quantum mechanics result. This is not
surprising since the definition of position operator is the same as
its definition in usual quantum mechanics.\\
In the same manner, a simple calculation gives,
\begin{equation}
\label{math:5.7}\langle
p\rangle=-i\sqrt{\frac{\hbar\omega}{2}}\Big[(\lambda-\lambda^{\ast})
-\sqrt{\frac{\hbar\omega\beta}{8}}[(\lambda-\lambda^{\ast})^2-1]\Big],
\end{equation}
and
$$\langle p\rangle^2=-\frac{\hbar\omega}{2}\Big\{(\lambda-\lambda^{\ast})^2
-2\sqrt{\frac{\hbar\omega\beta}{8}}(\lambda-\lambda^{\ast})[(\lambda-
\lambda^{\ast})^2-1]+$$
\begin{equation}
\label{math:5.8}
\frac{\hbar\omega\beta}{8}[(\lambda-\lambda^{\ast})^2-1]^2\Big\}.
\end{equation}
Since,
\begin{equation}
\label{math:5.9}p^2=-\frac{\hbar\omega}{2}\Big[(a_{\vec{k}}-a^{\dagger}_{\vec{k}})^2
-2\sqrt{\frac{\hbar\omega\beta}{8}}(a_{\vec{k}}-a^{\dagger}_{\vec{k}})^3+\frac{\hbar
\omega\beta}{8}(a_{\vec{k}}-a^{\dagger}_{\vec{k}})^4\Big],
\end{equation}
then,
\begin{equation}
\label{math:5.10} \langle
p^2\rangle=-\frac{\hbar\omega}{2}\Big\{[(\lambda-\lambda^{\ast})^2-1]-$$
$$2\sqrt{\frac{\hbar\omega\beta}{8}}(\lambda^3-{\lambda^{\ast}}^3-
3\lambda^{\ast}\lambda^2+3{\lambda^{\ast}}^2\lambda+3\lambda^{\ast}-3\lambda)+$$
$$\frac{\hbar\omega\beta}{8}(\lambda^4+{\lambda^{\ast}}^4-4\lambda^{\ast}\lambda^3-
4{\lambda^{\ast}}^3\lambda+6{\lambda^{\ast}}^2\lambda^2-6\lambda^2-6{\lambda^{\ast}}^2+
12\lambda^{\ast}\lambda+3)\Big\},
\end{equation}
and therefore one finds,
\begin{equation}
\label{math:5.11}(\Delta p)^2=\langle p^2\rangle-\langle
p\rangle^2=-\frac{\hbar\omega}{2}[-1-2\sqrt{\frac{\hbar\omega\beta}{8}}
(2\lambda^{\ast}-2\lambda)+\frac{\hbar\omega\beta}{8}(-4\lambda^2-4{\lambda^{\ast}}^2
+8\lambda^{\ast}\lambda+2)],
\end{equation}
or by some manipulations, one obtains the following result for
variance of $p$,
\begin{equation}
\label{math:5.12}({\Delta
p})^2=\frac{\hbar\omega}{2}+\hbar\omega\sqrt{\frac{\hbar\omega\beta}{2}}
(\lambda^{\ast}-\lambda)+\frac{\hbar^2\omega^2\beta}{8}[1-2(\lambda^{\ast}-\lambda)^2]
\end{equation}
Note that these results give the usual quantum mechanical results
when $\beta \longrightarrow 0$. Equations (45) and (51) show that
although the definition of coherent states do not changes in GUP,
but because of quantum gravitational effect expectation values and
variances change considerably. Now product $({\Delta p})^2({\Delta
x})^2$ has a complicated form which shows that complete coherency is
impossible. Therefore, there is a considerable departure from very
notion of coherency. In usual quantum mechanics one can have
complete coherency in principle. One can localize wave packet in
space completely, at least in principle, and wave can propagate
without broadening, at least in principle. This is evident from
$\Delta x\geq\frac{\hbar}{\Delta p}$. In quantum gravity because of
gravitational induced uncertainty, one can not localize wave packet
at all and it is impossible to cancel out broadening. Therefore in
quantum gravity one can not have any solitonic states and any wave
packet will suffer more broadening.
\section{Wave Packet Propagation}
The problem of wave packet propagation in quantum gravity first has
been considered by Amelino-Camelia {\it et al}[27]. Using a
$\kappa$-deformed Minkowski spacetime, they have investigated the
experimental testability concerning the $\kappa$-deformed Minkowski
relation between group velocity and momentum. Amelino-Camelia and
Majid have considered the problem of waves propagation in
Noncommutative Spacetime[28]. They have considered quantum group
Fourier transform methods applied to the study of processes on
noncommutative Minkowski spacetime. They have derived a wave
equation  and have investigated the associated phenomena of in vacuo
dispersion. Assuming the deformation scale to be of the order of the
Planck length they have found that the dispersion effects are large
enough to be tested in experimental investigations of astrophysical
phenomena such as gamma-ray bursts. Here in a simpler approach, we
will show that there is an additional broadening for wave packet due
to gravitational induced uncertainty. This can be considered as a
result of generalized dispersion relations or due to the variations
in universal constants.

\subsection{ Wave Packet Propagation in Ordinary Quantum Mechanics}
Consider the following plane wave profile,
\begin{equation}
\label{math:1.1} f(x,t)\propto e^{\, ikx-i\omega t}.
\end{equation}
Since  $\omega= 2\pi\nu$,  $k=\frac{2\pi}{\lambda}$  and
$\nu=\frac{c}{\lambda}$,  this equation can be written as
$f(x,t)\propto e^{\, ik(x-ct)}$. Now the superposition of these
plane waves with amplitude $g(k)$ can be written as,
\begin{equation}
\label{math:1.2}f(x,t)=\int_{-\infty}^\infty dk\, g(k)\, e^{\,
ik(x-ct)}=f(x-ct)
\end{equation}
where $g(k)$ can have Gaussian profile. This wave packet is
localized at $x-ct=0$. In the absence dispersion properties for the
medium, wave packet will not suffers any broadening with time. In
this case the relation $\omega=kc$ holds. In general the medium has
dispersion properties and therefore $\omega$ becomes a function of
wave number, $\omega=\omega(k)$. In this situation equation (53)
becomes,
\begin{equation}
\label{math:1.3}f(x,t)=\int dk\, g(k)\, e^{\, ikx-i\omega(k)t}.
\end{equation}
Suppose that $g(k)=e^{-\alpha(k-k_{0})^{2}}$. With expansion of
$\omega(k)$ around $k=k_{0}$, one find
\begin{equation}
\label{math:1.4}\omega(k)\approx
\omega(k_0)+(k-k_0)\bigg({\frac{d\omega}{dk}}\bigg)_{k_0}+\frac{1}{2}(k-k_0)^2\bigg({\frac{d^2\omega}{dk^2}}\bigg)_{k_0},
\end{equation}
where using the definitions,
\begin{equation}
\label{math:1.5}
\bigg({\frac{d\omega}{dk}}\bigg)_{k_0}=v_g,\quad\quad\frac{1}{2}\bigg({\frac{d^2\omega}{dk^2}}\bigg)_{k_0}=\mu
,\quad\quad k-k_0=k^\prime.
\end{equation}
equation (54) can be written as,
$$f(x,t)=e^{\,ik_0x-i\omega(k_0)t}\int_{-\infty}^\infty dk^\prime
\,e^{-\alpha{k^\prime}^2}e^{\,ik^\prime(x-v_gt)}
\,e^{-i{k^\prime}^2\beta t}$$
\begin{equation}
\label{math:1.6} =e^{\,ik_0x-i\omega(k_0)t}\int_{-\infty}^\infty
dk^\prime \,e^{\,ik^\prime(x-v_gt)}\,e^{-(\alpha+i\mu
t){k^\prime}^2}.
\end{equation}
Now completing the square root in exponent and integration gives,
\begin{equation}
\label{math:1.7}f(x,t)=e^{\,i\big[k_0x-\omega(k_0)t\big]}\,\bigg(\frac{\pi}{\alpha+i\mu
t}\bigg)^{\frac{1}{2}}\,e^{-\big[\frac{(x-v_gt)^2}{4(\alpha+i\mu
t)}\big]}.
\end{equation}
Therefore one find,
\begin{equation}
\label{math:1.8}|f(x,t)|^2=\bigg(\frac{\pi^2}{\alpha^2+\mu^2
t^2}\bigg)^{\frac{1}{2}}e^{-\big[\frac{\alpha(x-v_gt)^2}
 {2(\alpha^2+\mu^2
t^2)}\big]},
\end{equation}
which is the profile of the wave in position space. The quantity
which in $t=0$ was $\alpha$, now has became $\alpha+\frac{\mu^2
t^2}{\alpha}$ and this is the notion of broadening. Therefore,
\begin{equation}
\label{math:1.9}Broadening\propto \bigg(1+\frac{\mu^2
t^2}{\alpha^2}\bigg)^{1\over2}.
\end{equation}
This relation shows that a wave packet with width $(\Delta x)_{0}$
in $t=0$ after propagation will have the following width,
\begin{equation}
\label{math:1.10}(\Delta x)_{t} = (\Delta x)_{0}\bigg(1+\frac{\mu^2
t^2}{\alpha^2}\bigg)^{1\over2}.
\end{equation}
\subsection{ Wave Packet Propagation in Quantum Gravity} As has been
indicated, when one considers gravitational effects, usual
uncertainty relation of Heisenberg should be replaced by,
\begin{equation}
\label{math:2.1}\Delta x\geq\frac{\hbar}{\Delta
p}+\frac{\alpha^\prime l_p^2\Delta p}{\hbar}.
\end{equation}
As a first step analysis we consider the above simple form of GUP.
Suppose that
$$\Delta x\sim x,\quad\quad\Delta p\sim p,\quad\quad p=\hbar k,\quad\quad x=\bar{\lambda}=\frac{\lambda}{2\pi}.$$
Therefore one can write,
\begin{equation}
\label{math:2.2}\bar{\lambda}=\frac{1}{k}+\alpha^\prime l_p^2\,k
\qquad and \qquad \omega=\frac{c}{\bar{\lambda}}.
\end{equation}
In this situation  the dispersion relation becomes,
\begin{equation}
\label{math:2.3}\omega=\omega(k)=\frac{kc}{1+\alpha^\prime
l_p^2\,k^2}.
\end{equation}
This relation can be described in another viewpoint. By expansion of
$\bigg(1+\alpha^\prime l_p^2\,k^2\bigg)^{-1}$ and neglecting second
and higher order terms of $\alpha^{\prime}$, we find that
$\omega=kc\big(1-\alpha^\prime l_p^2\,k^2\big)$. This can be
considered as $\omega=k^{\prime}c$ where
$k^{\prime}=k\big(1-\alpha^\prime l_p^2\,k^2\big)$. Now one can
define a generalized momentum as $p=\hbar k^{\prime}=\hbar
k\big(1-\alpha^\prime l_p^2\,k^2\big)$. It is possible to consider
this equation as $p=\hbar^{\prime}k$ where
$\hbar^{\prime}=\hbar\big(1-\alpha^\prime l_p^2\,k^2\big)$. So one
can interpret it as a wave number dependent Planck "constant". In
the same manner group velocity becomes,
\begin{equation}
\label{math:2.4}v_g=\frac{d\omega}{dk}\bigg|_{k=k_0}=\frac{c(1-\alpha^\prime
l_p^2\,k^2)}{(1+\alpha^\prime l_p^2\,k^2)^2}\bigg|_{k=k_0}.
\end{equation}
Up to first order in $\alpha^{\prime}$ this relation reduces to
$v_{g}\approx c\big(1-3\alpha^{\prime}l_{p}^{2}k_{0}^{2}\big)$.\\
A little algebra gives $\mu$ as follow
\begin{equation}
\label{math:2.5}\mu=\frac{1}{2}\bigg(\frac{d^2\omega}{dk^2}\bigg)\bigg|_{k=k_0}
=\frac{-3\alpha^\prime l_p^2c\,k(1+\alpha^\prime l_p^2
\,k^2)^2+4{\alpha^\prime}^2l_p^4c\,k^3(1+\alpha^\prime l_p^2
\,k^2)}{(1+\alpha^\prime l_p^2\,k^2)^4}\Bigg|_{k=k_0},
\end{equation}
which up to first order in $\alpha^{\prime}$ reduces to
 $\mu \approx-3\alpha^\prime l_p^2ck_{0}$. It is evident that when
$\alpha^\prime\rightarrow0$ then $\mu \rightarrow 0$ and
$v_g\rightarrow c$. The same analysis which has leads us to equation
(61), now gives the following result,
\begin{equation}
\label{math:2.6}(\Delta x)_{t} = (\Delta
x)_{0}\Bigg(1+\frac{1}{\alpha^2}\bigg(\frac{-3\alpha^\prime
l_p^2c\,k_{0}(1+\alpha^\prime l_p^2
\,k_{0}^2)^2+4{\alpha^\prime}^2l_p^4c\,k_{0}^3(1+\alpha^\prime l_p^2
\,k_{0}^2)}{(1+\alpha^\prime l_p^2\,k_{0}^2)^4}\bigg)^2
t^2\Bigg)^{1\over2}.
\end{equation}
If one accepts that $\alpha^{\prime}$ is negative constant (
$\alpha^{\prime}<0$), then group velocity of the wave packet becomes
greater than light velocity. This is evident from equation (65) and
is reasonable from varying speed of light models. In fact if
$|\alpha^{\prime}|k^{2}l^{2}_{p}\ll1$, one recover usual quantum
mechanics but when $|\alpha^{\prime}|k^{2}l^{2}_{p}\approx1$, Planck
scale quantum mechanics will be achieved. Based on this argument,
equation (67) shows that in quantum gravity there exists a more
broadening of wave packet due to gravitational effects. Up to first
order in $\alpha^{\prime}$, this equation becomes,
\begin{equation}
\label{math:2.6}(\Delta x)_{t} = (\Delta
x)_{0}\Bigg(1-\frac{3\alpha^\prime
l_p^2c\,k_{0}t^2}{\alpha^2}\Bigg)^{1\over2}.
\end{equation}
Now using equation (64), one can write the generalized dispersion
relation as the following form also,
\begin{equation}
\label{math:2.7}\omega(p)=\frac{\hbar pc}{\hbar^2+\alpha^\prime
l_p^2\,p^2},
\end{equation}
or
\begin{equation}
\label{math:2.8}E^\prime=\hbar \omega(p)=\frac{p
c}{1+\alpha^\prime\Big(\frac{l_pp}{\hbar}\Big)^2}.
\end{equation}
It is evident that if $\alpha^\prime\longrightarrow0$ Then
$E^\prime\longrightarrow E=pc$  and
$\omega(p)\longrightarrow\omega=\frac{pc}{\hbar}$.

\section{Summary}
In this paper the effect of gravitation on some well-known quantum
optical phenomena has been studied. Considering dynamics and quantum
mechanical coherent states of a simple harmonic oscillator in the
framework of Generalized Uncertainty Principle(GUP), we have derived
the equation of motion for simple harmonic oscillator and some of
their new implications have been discussed. As an important
consequence we have shown that essentially, there is no harmonic
oscillation in quantum gravity regime. Then coherent states of
harmonic oscillator in the case of GUP are compared with relative
situation in ordinary quantum mechanics. It is shown that in the
framework of GUP there is no considerable difference in definition
of coherent states relative to ordinary quantum mechanics. But,
considering expectation values and variance of some operators, based
on quantum gravitational arguments one concludes that although it is
possible to have complete coherency and vanishing broadening in
usual quantum mechanics, gravitational induced uncertainty destroys
complete coherency in quantum gravity and it is impossible to have a
monochromatic ray in principle. Finally we have shown that there is
an extra broadening in wave packet propagation due to quantum
gravitational effects. This leads us to generalized dispersion
relation. Generalized dispersion relations can be described as a
possible framework for varying constant of the nature. Since quantum
gravitational effects are very small, their possible detection
requires very high energy experiments. It seems that LHC will
provide a reasonable framework for testing these predictions.

\end{document}